\documentclass[11pt]{article}
\usepackage{moriond,epsfig}
\input babarsym

\def\allsemil{  32210 \pm    230}

\def\allbrbr{0.0197 \pm 0.0025} 
\def\allbrbrerrmc{0.0010}

\def\bchbrbr{0.0168 \pm 0.0030}

\def\bnebrbr{0.0246 \pm 0.0043}

\def\allbr{ 2.14}
\def\allbrE{ 0.29}
\def\allbrEsyst{ 0.26}
\def\allbrEthHi{ 0.37}

\def\allvub{ 4.52}
\def\allvubE{ 0.31}
\def\allvubEsyst{ 0.27}
\def\allvubEthHi{ 0.40}

\def\allvubEpert{ 0.09}
\def\allvubEhomb{ 0.24}

\def\ra{\rightarrow}

\def\be{\begin{equation}}
\def\ee{\end{equation}}
\def\bea{\begin{eqnarray}}
\def\eea{\end{eqnarray}}

\begin{document}
\begin{flushright}
SLAC-PUB-9922 \\
\babar-PROC-03-011 \\
June 2003 \\
\end{flushright}
\vspace*{4cm}
\title{ RECENT RESULTS ON \mathversion{bold}$|V_{ub}|$, \mathversion{bold}$|V_{cb}|$ AND MIXING FROM {\sc BaBar}}

\author{ D. del Re \\
	(representing the {\sc BaBar} Collaboration)}

\address{Universit\`a di Fisica ``La Sapienza'', Dipartimento di Fisica, P.le Aldo Moro 2, Rome, Italy \\
	and\\
	University of California, San Diego, Department of Physics, La Jolla, CA 92093, USA}

\maketitle\abstracts{We present the measurements of the CKM matrix parameters
$V_{ub}$, $V_{cb}$ and of the $B$ mixing oscillation frequency with the 
{\sc BaBar} experiment
at the asymmetric $B$-factory PEPII.
Data were collected in the years 2000-2002
and the total available statistics corresponds to 91 $fb^{-1}$.
The $V_{ub}$, $V_{cb}$ measurements utilize both inclusive and 
exclusive semileptonic decays of the $B$ meson.
The $\Delta m_d$ parameter is measured by using the time evolution of 
the $B$, determined from the flight length difference between the two 
$B$ mesons. }

\section{Introduction}

Precise measurements of the matrix elements \Vub and \Vcb 
and the mixing $\Delta m_d$ parameter allow to perform tests of the unitarity of Cabibbo-Kobayashi-Maskawa (CKM) 
matrix \cite{ckm}. 
In this paper I present a measurement of \Vub using inclusive semileptonic \B decays.
The analysis is based on \BB\ events in which one of  the $B$ mesons decays hadronically and is
fully  reconstructed ($B_{reco}$)  and the  semileptonic decay  of the
recoiling  $B$   meson  ($B_{recoil}$)   is  detected  by   a  charged
lepton. To separate the signal decays \Bxulnu\ from the background we use 
the invariant  mass $\mX$ of the
hadronic system recoiling against the charged lepton and the neutrino.
Moreover I present a measurement of the exclusive semileptonic branching ratio $\BR (\BztoDslnu )$.
This channel is of particular physical interest, since the study of its differential decay rate can be 
used to measure the element \Vcb.
Finally the sample of \BztoDslnu decays has been also used to extract the $\Delta m_d$ mixing 
parameter.

\subsection{The Dataset and the \babar\ Experiment}\label{subsec:dataset}

These measurements are based on data recorded by the \babar\
detector~\cite{babarnim} at the \pep2\ asymmetric-energy \epem storage
ring at SLAC. The data sample of about 88 million \BB\ pairs
corresponds to an integrated luminosity of 81.9\invfb collected at the
\FourS\ resonance.  The \babar\ detector consists of a five-layer
silicon vertex tracker (SVT), a 40-layer drift chamber (DCH), a
detector of internally reflected Cherenkov light (DIRC), an
electromagnetic calorimeter (EMC), assembled from 6580 CsI(Tl)
crystals, all embedded in a solenoidal magnetic field of $1.5\,{\rm T}$ and
surrounded by an instrumented flux return (IFR).

\section{Measurement of the Inclusive  Charmless Semileptonic
Branching Fraction of $B$ Mesons and Determination of \mathversion{bold}$|V_{ub}|$}

\subsection{Signal Simulation}
In  charmless  semileptonic  decays,  we  expect  resonant  states  to
dominate  at  low  recoil  masses,   while  at  high  mass  we  expect
non-resonant multi-hadron states  to prevail.  Consequently, we choose
to employ  a hybrid model.  We use  the ISGW2 model~\cite{ref:isgwtwo}
to produce the pseudo-scalar,  vector, and heavier mesons to represent
25\%, 35\%,  and 40\%, of  the resonant states, respectively;  this is
consistent with currently measured branching fractions~\cite{pdg2002}.
We  add  a  non-resonant  component  on  the basis  of  HQET  and  OPE
calculations~\cite{ref:fazioneubert},   where  the   fragmentation  is
handled by  JETSET~\cite{ref:jetset}. The two  components are combined
so  that the  inclusive charmless  semileptonic branching  fraction is
compatible with this  measurement and the LEP average  and so that the
integral distribution function is consistent with the OPE calculation,
except for  discontinuities (due to resonances) at  the lowest masses.
The  non-resonant  component  amounts  to  about  66\%  of  the  total
charmless  semileptonic branching  fraction.   The motion  of the  $b$
quark  inside the $B$  meson is  implemented with  the parametrization
described in~\cite{ref:fazioneubert}.

\subsection{Selection of Hadronic $B$ Decays, $B \ra \Db Y$}
This analysis is based on \BB\ events in which one of  the $B$ mesons 
decays hadronically.
Hadronic $B$  decays of the type  $B \rightarrow \Db  Y$ are selected,
where $D$ refers to a charm  meson, and $Y$ represents a collection of
hadrons with  a total  charge of $\pm  1$, composed  of $n_1\pi^{\pm}+
n_2K^{\pm}+ n_3\KS +  n_4\piz$, where $n_1 + n_2 < 6$,  $n_3 < 3$, and
$n_4 <  3$.  Using $D^-$  and $D^{\star-}$ ($\Dzb$ and  $\Dstarzb$) as
seeds for  $B^0$ ($B^+$) decays,  we reconstruct about  1000 different
decay chains.
The efficiency  to reconstruct a  hadronic $B$ decay
amounts to $0.15\%$ for \Bz\ mesons and to $0.25\%$ for \Bp\ mesons. 

The kinematic consistency of a $B_{reco}$ candidate with a $B$ meson
decay is checked using two variables, the beam energy-substituted mass
$\mes = \sqrt{s/4 - \vec{p}^{\,2}_B}$
 and the energy difference, 
$\Delta E  = E_B  - \sqrt{s}/2$. Here  $\sqrt{s}$ refers to  the total
energy in the \FourS center of  mass frame, and $p_B$ and $E_B$ denote
the momentum and energy of the $B_{reco}$ candidate in the same frame,
respectively.  We  require $\Delta E  = 0$ within  approximately three
standard deviations. 

\subsection{Selection of Semileptonic Decays, $\Bxlnu$}

Semileptonic $B$ decays are identified  by the presence of an electron
or  muon candidate.   To reduce  backgrounds from  secondary  charm or
$\tau^{\pm}$ decays and from fake  leptons, we only retain events with
a charged  lepton with  a minimum  momentum in the  $B$ rest  frame of
$p^*_{\ell} > 1 \gevc$. The details of both the lepton identifications are described elsewhere \cite{babarnim}.

To further improve  this rejection, we require that  the lepton charge
and the  flavor of the  reconstructed $B^{\pm}$ are consistent  with a
prompt semileptonic  decay of the  $B_{recoil}$.  For $B^0$ we  do not
impose this  restriction, instead we take into  account \Bz-\Bzb\ mixing
and correct for the contribution from secondary leptons.

\subsection{Selection of $\Bxulnu$ Decays}
The hadron system  $X$ in the decay $\Bxlnu$ is constituted
from  charged tracks  (excluding  leptons with  $p^* >  1 \gevc$),
photons,  and  $K^0$  that  are  not associated  with  the  $B_{reco}$
candidate.   Specifically, we  select charged  tracks in  the fiducial
volume $0.41 < \theta_{lab} <  2.54\rad$. Photons are identified as energy
clusters  in the  EMC that  are spatially  separated from  any charged
track, have an energy $E_{lab}>80
\mev$, and have a lateral cluster shape consistent with a photon. Care
is taken to eliminate duplicate,  curling, and fake charged tracks, as
well as low energy photons from beam background and energy depositions
in the EMC by charged and neutral hadrons.

We improve the resolution in the measurement of $\mX$ with a 2-C
kinematic fit that imposes four-momentum conservation, the equality of
the masses of the two $B$ mesons and forces
$p_{\nu}^2 = 0$.  

The selection of $\Bxulnu$ decays is tightened by requiring
(1) one and only one charged lepton with $p^*_{\ell} > 1 \gevc$ in the
    event, 
(2) charge conservation, \ie,  the total charge of the $B_{reco}$,
    electron and the hadron system is required to be $Q_{tot} = 0$; and
(3)   for the missing mass of the event,  $\mmiss < 0.5 \gev^2/c^4$.
These criteria not  only improve the resolution in  the measurement of
the missing neutrino, they  also suppress the dominant \Bxclnu\ decays,
many of  which contain additional  neutrinos or undetected  $K_L$.

To further  suppress $\Bzb  \to\Dstarp\ell^-\overline{\nu}$  decays, we
exploit  the  fact that  the  momentum of  the  pion  produced in  the
$\Dstarp\to\Dz\pi^+$  decay  is  almost  collinear with  the  $D^{*+}$
momentum in  the laboratory frame.   In that frame we  can approximate
the energy of the $D^{*+}$ as $E_{\Dstarp} \simeq
\gamma m_{\Dstarp}= E_\pi \cdot m_{\Dstarp}/145 \mev$ and then require for the
mass of the neutrino $ m_{\nu}^2 = (p_{B_{recoil}}-p_{D^*} - p_{\ell})^2 <
- 3(\gevcc)^2$. 
This selection removes about 68\% of the $\Bzb\to \Dstarp\ell^- \bar{\nu}$ background
without significantly affecting the signal $\Bxulnu$
decays.

The detection of kaons in the recoil system is a powerful tool to veto
background,  since  less than  $\sim  10\%$  of  all $\Bxulnu$  decays
contain  kaons, whereas  a  large fraction  of  charm particles decay
in charged or neutral kaons.
Charged kaons are identified by an algorithm 
based on the DIRC, DCH and the SVT information.
$K_S\to\pi^+\pi^-$ decays are selected from pairs
of oppositely charged tracks that are kinematically constrained to
originate from a common vertex, and have an invariant mass consistent
with the $K_S$ mass, $0.486 < m_{\pi^+\pi^-} < 0.510 \gevcc$.

We finally separate the events into two subsamples, signal enriched and signal depleted
sample, depending on the absence or presence of at  least one
charged or neutral kaon in  the recoiling system.  The depleted sample
is used as a control sample to study  in detail  that the Monte  Carlo  simulation
reproduces the resolutions accurately.

\subsection{Extraction of Branching Ratios}
\label{sec:fitconcepts}

We  derive $|V_{ub}|$  from a  measurement of  the ratio  of branching
fractions 

\begin{equation}
\rusl=
\frac{\BR(\Bxulnu)}{\BR(\Bxlnu)}=
\frac{N_u/(\varepsilon_{sel}^u \varepsilon_{\mX}^u)}{N_{sl}} 
\times \frac{\varepsilon_l^{sl} \varepsilon_t^{sl} } {\varepsilon_l^u \varepsilon_t^u },
\label{eq:vubExtr}
\end{equation}

\noindent observed for the admixture of \Bz\ and \Bp\  in the \breco\
tagged sample. 

   \begin{figure}
    \begin{centering}
    \epsfig{file=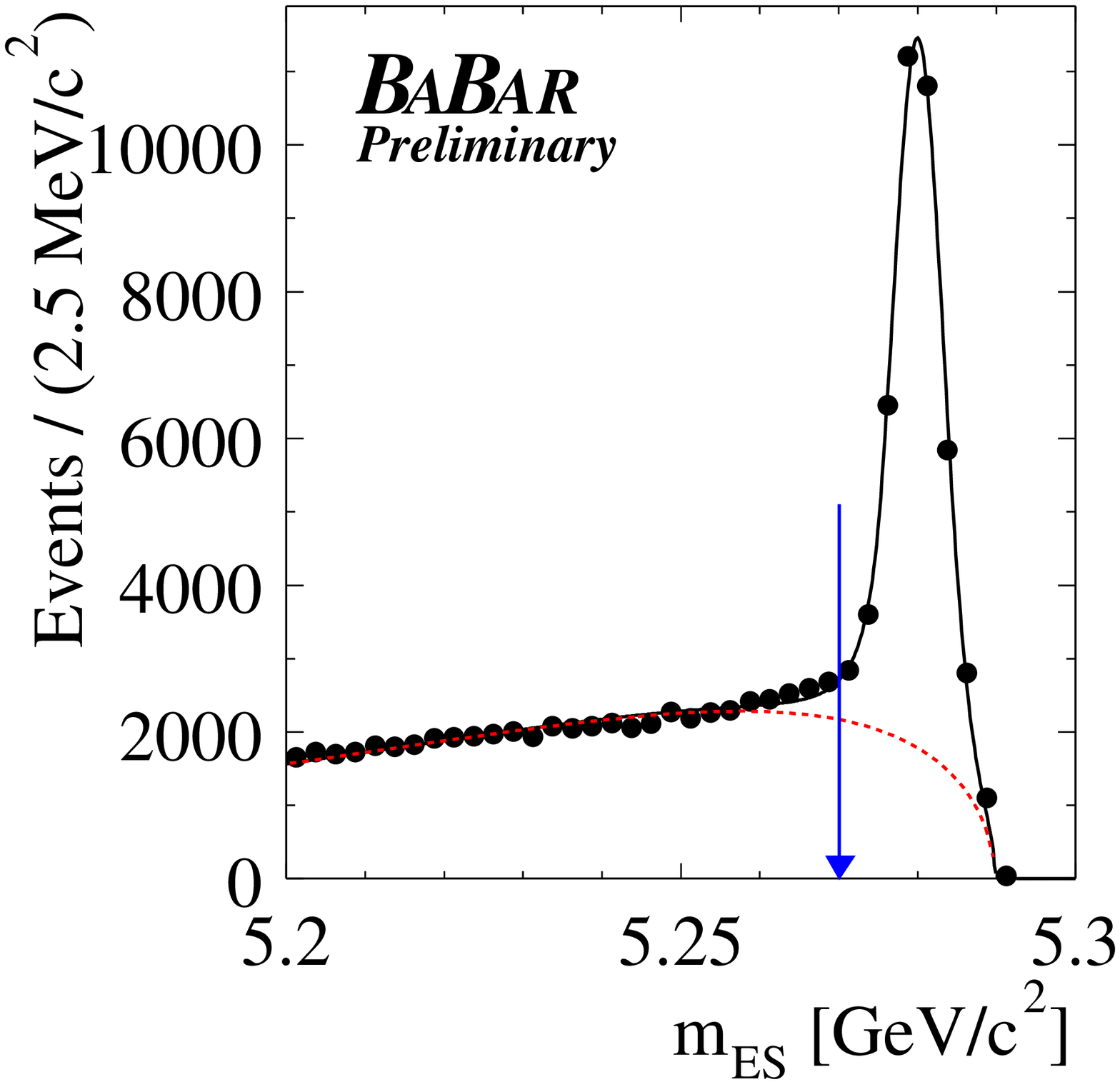,height=6.0cm}  \epsfig{file=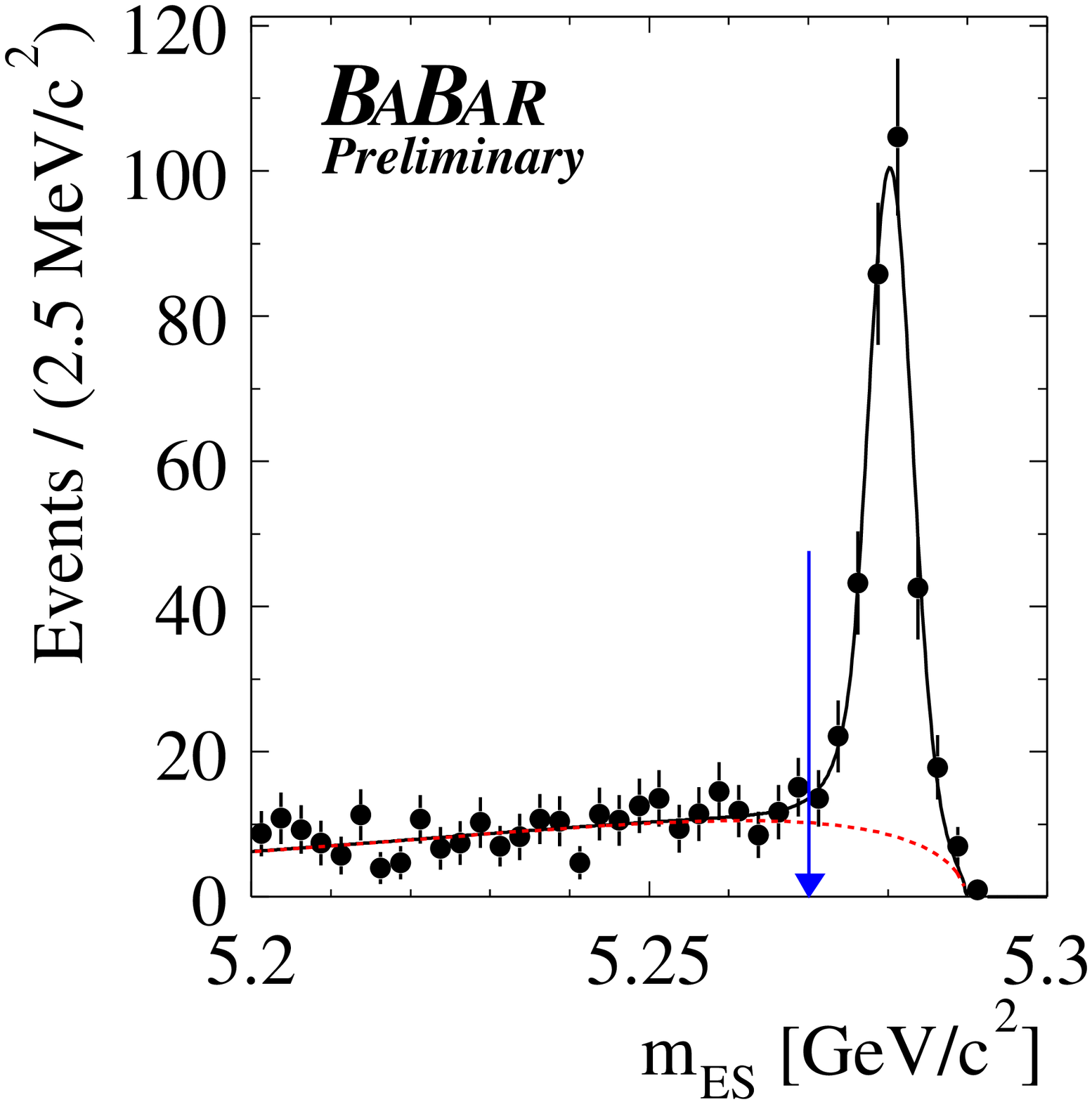,height=6.0cm} 

    \caption{Fit to the $\mes$ distributions for the  lepton
    sample with $p^* > 1\gevc$ in the recoil of a \breco\ candidate
    (left), and for selected  events with $\mX < 1.55 \gevcc$
    (right). The arrow indicates the cut on \mes.
   \label{fig:msl}}
   \end{centering}
   \end{figure}
The denominator $N_{sl}=N_{sl}^{meas} - BG_{sl}$ is the number of
events with at least one charged lepton fulfilling the charge-flavor
correlation.  $N_{sl}^{meas}$ is derived from a fit to the $\mes$
distribution with the sum of an empirical background
function~\cite{argusf} describing the contribution from both continuum
events and combinatorial background from $\BB$ events and a narrow
signal~\cite{cry} peaked at the $B$ meson mass. 
This fit, shown in Figure ~\ref{fig:msl}a, results in $N_{sl} =
\allsemil$ events.  The residual background $BG_{sl}$ is mainly due to
misidentified leptons and semileptonic charm decays; it is estimated
by Monte Carlo simulation to be $(6.8
\pm 1.2)\%$ of the total lepton yield.

The signal  yield $N_u^{meas}= N_u  + BG_u$ is determined  from events
with $\mX < m_X^{cut}$ in  the signal enriched sample (veto on kaons).
In each bin of the \mX\ distribution, the combinatorial
$B_{reco}$ background is subtracted based on a fit to the $\mes$
distribution.  Figure~\ref{fig:msl}b shows the $\mes$ distribution for
the low mass region $\mX<1.55 \gevcc$; we require $\mes > 5.27
\gevcc$.  

\begin{figure}
    \begin{centering}
     \epsfig{file=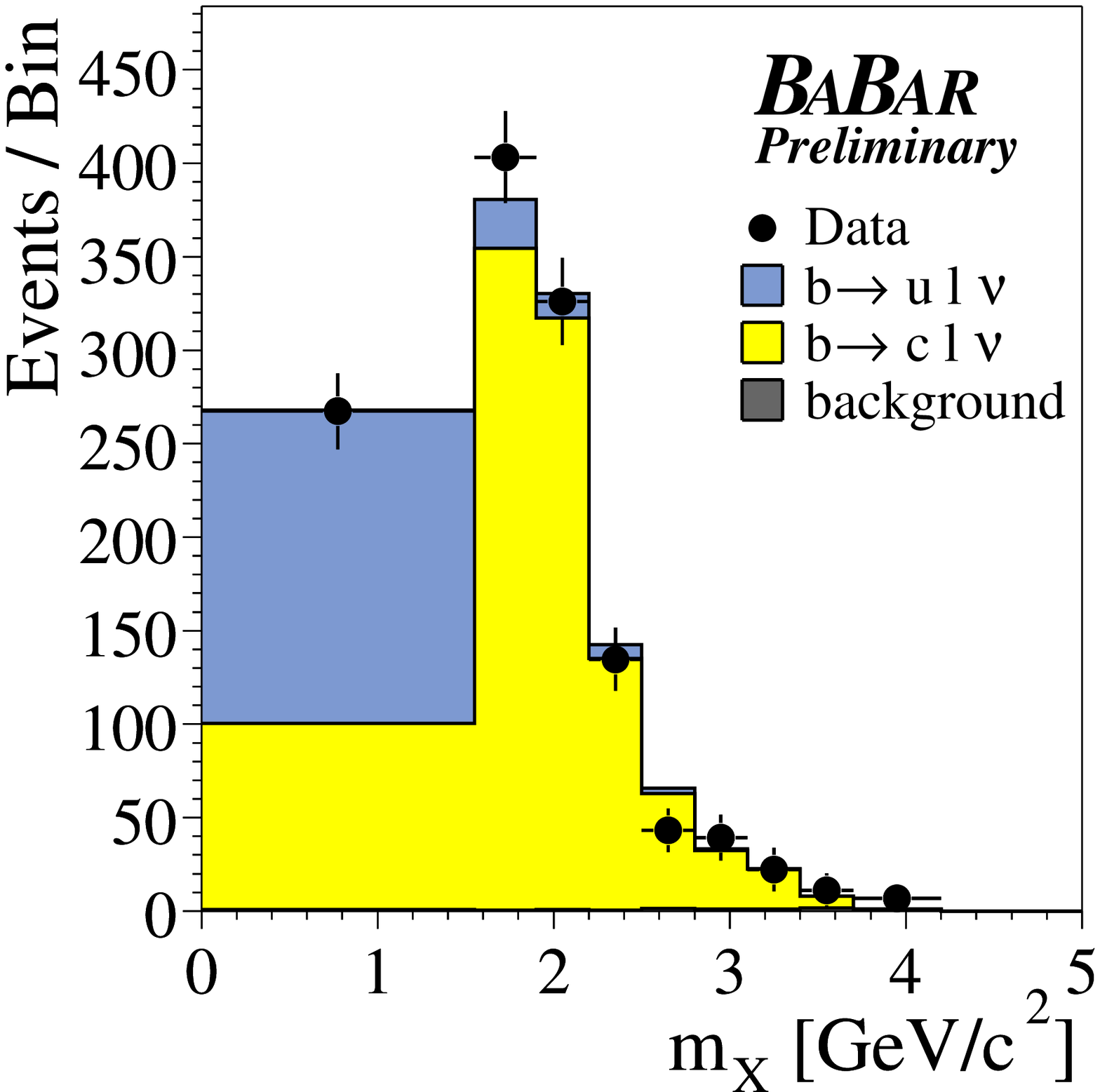,height=6.0cm} 
     \epsfig{file=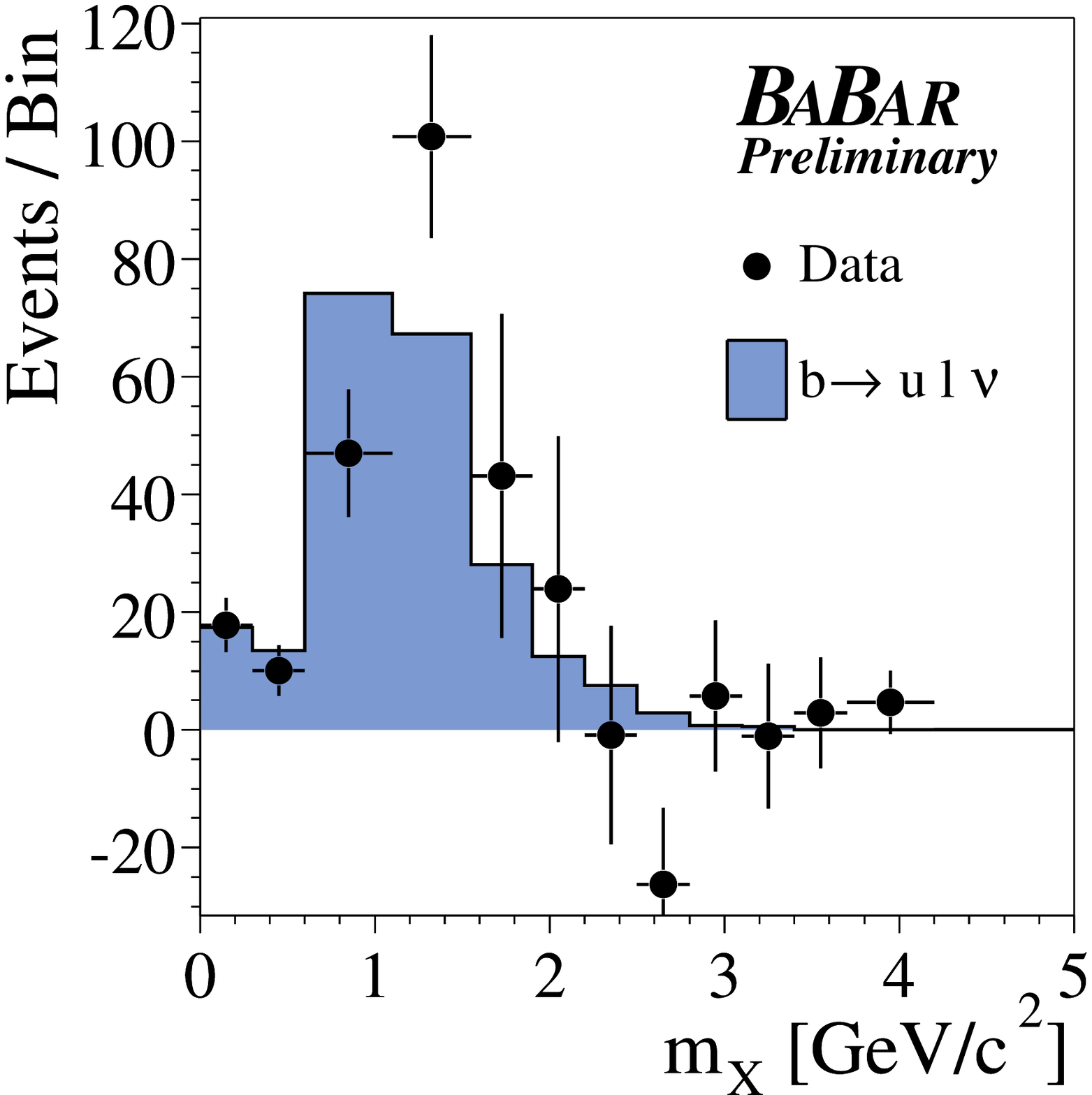,height=6.0cm}
     \caption{A $\chi^2$ fit to the $\mX$ distribution for the $\Bxulnu$
     enriched sample. Left: the data (points) and three fitted
     contributions. Right: background subtracted data and signal MC.
   \label{fig:fitdata}}
   \end{centering}
   \end{figure}

The number of signal events $N_u$ is extracted from a binned $\chi^2$
fit of the measured $\mX$ distribution to the sum of three
contributions: the signal, the background from $\Bxclnu$,
and the background from other sources (misidentified leptons,
secondary $\tau$ and charm decays).  The shape of the signal and
background contributions are derived from Monte Carlo simulation and
their relative normalizations are free parameters in the fit.  The
$\chi^2$ function takes into account the statistical errors from the
$\mes$ fit and from the Monte Carlo sample.
Figure~\ref{fig:fitdata} shows  the $\mX$ distribution  (corrected for
$B_{reco}$ background) for the $b  \ra u$ enriched sample.  Also shown
are  the results  of  the fit,  with  the fitted  signal  and the  two
background   distributions.   The  fit   reproduces  the   data  well,
$\chi^2/dof=0.85$.  

The following efficiency corrections have been determined from Monte
Carlo simulations: $\varepsilon^u_{sel}$ is the efficiency for
selecting $\Bxulnu$ decays with all analysis requirements and is
found to be $\varepsilon^u_{sel}=0.342\pm0.006(stat)$.
$\varepsilon^u_{\mX}$ is the efficiency of the cut $m_X^{cut}$ and
depends on the theoretical parameters of the Fermi motion.  For
$m_X^{cut}=1.55\gevcc$, we obtain
$\varepsilon^u_{\mX}=0.733\pm0.009(stat)$.

The following ratios are determined from Monte Carlo simulations:
the factor $\varepsilon_l^{sl}/\varepsilon_l^u $ corrects for the difference
of the impact of the lepton momentum cut for $\Bxlnu$
and  $\Bxulnu$  decays.  We  estimate for  electrons
$\varepsilon_e^{sl}/\varepsilon_e^{u}=0.896\pm 0.009$,     for     muons
$\varepsilon_{\mu}^{sl}/\varepsilon_{\mu}^{u}=0.875\pm  0.015$,  and for  the
combined  sample  $\varepsilon_{\ell}^{sl}/\varepsilon_{\ell}^u  = 0.887\pm0.008$.
The ratio $\varepsilon_t^{sl}/\varepsilon_t^u $ accounts for a possible
difference in the efficiencies for finding a $B_{reco}$ decay in
events with semileptonic decays in the recoiling $B$, $\Bxlnu$ and
$\Bxulnu$.  This ratio is expected to be close to one.  Based on
Monte Carlo simulations, we find $\varepsilon_t^{sl}/\varepsilon_t^u = 0.98
\pm 0.04$. We thus assume $\varepsilon_t^{sl}/\varepsilon_t^u = 1.00 \pm
0.04$ and take the statistical  error of the Monte Carlo simulation as
systematic error.

With these corrections, we obtain for the ratio of branching fractions
\begin{equation}
\rusl= 
\frac{\BR(\Bxulnu)}{\BR(\Bxlnu)}
= \allbrbr (stat)  \pm \allbrbrerrmc (MC\, stat)
\end{equation}

We performed the measurement on various subsamples of data.
For \Bz we measure $\rusl=\bnebrbr(stat)$ and for
\Bp we  obtain $\rusl  = \bchbrbr(stat)$. The  result is also stable for
different  choices   of  $\mX^{cut}$,  resulting   in  widely  varying
signal/background ratios. 

\subsection{Systematic Uncertainties}
\label{sec:systematics}

Extensive  studies  have  been  performed  to  assess  the  systematic
uncertainties and  potential biases that might  impact the measurement
of  $\rusl$.  

The reconstruction  of the  invariant mass of  the hadronic  recoil is
affected  by  missing  and  fake  particles,  as  well  as  resolution
effects. Taking into account the uncertainties related to the 
reconstruction of charged tracks and neutral  particles (including 
the uncertainty in the energy deposition of $K_L$ in  the calorimeter)  
and to particle  identification by the usage of indipendent control samples we get 
a relative systematics uncertainty on the branching ratio of $7.5\%$.

The uncertainty of the $B_{reco}$ combinatorial background subtraction
is estimated by changing the signal shape function to a Gaussian and
by varying the parameters within one standard deviation of the default
values.  The binning systematics have been studied by varying the
binning for $\mX > 1.55\gevcc$. The resulting systematics uncertainty
is $6.2\%$.

While  the  total semileptonic  branching  fraction  $\Bxlnu$ is  well
measured, the  decomposition to individual  charm states is  much less
certain.  The impact on $\rusl$  has been estimated by varying each of
the exclusive  branching ratios within  one standard deviation  of the
current world average~\cite{pdg2002}.  Similarly, the branching ratios
of charm mesons have been varied, both for exclusive decays and
for inclusive kaon production. The variation on the branching ratio 
is $4.4\%$.

The  signal detection efficiency  $\varepsilon^u_{sel}$ depends  on event
selection  criteria, the composition  and average  mass of  the hadron
system and the decay of the final state particles. We study the impact
of  signal events  with two  kaons by  including events  with  an even
number of kaons in the signal enriched sample.  The uncertainty in the
hadronization of the  final state of \Bxulnu\ events  is determined by
measuring $\rusl$ in bins  of charged and neutral multiplicities.  The
result agrees  well with the  standard fit, the difference  is smaller
than  $1\%$.   In  addition,  we   perform  the  fit  using  only  the
non-resonant signal model instead of the hybrid model.  The difference
to the default method is $3\%$,  which we take as an estimate of the
systematic error  for the hadronization uncertainties.   We vary the
inclusive and exclusive branching   fractions  for   the   charmless   semileptonic
$B$-decays  by  one  standard  deviation. The fraction of \ssbar\ 
popping events is varied by $100\%$ in the 
exclusive signal contribution and by $30\%$ in the inclusive one. In summary
the uncertainty on the \Bxulnu\ modeling gives a systematic effect of $4.1\%$.

Due  to the  large  difference in  the  reconstruction efficiency  for
charged  and neutral  $B_{reco}$  candidates, the  event sample  contains
substantially more \Bp\ than \Bz  mesons. This has a negligible impact
on the extraction of  the charmless semileptonic branching fraction and
\Vub. No correction is applied. 

The fraction $\varepsilon_{\mX}$ of signal events with $\mX < 1.55
\gevcc$ is taken from a  theoretical calculation of the shape of the
hadronic  recoil  mass distribution  and  depends on  non-perturbative
parameters:   $\lbar  =   0.480   \pm  0.122   \gev$   and  $\lone   =
-0.300\pm0.105\gev^2$,  where we  have taken  the recent  measurement of
CLEO~\cite{ref:cleomxhad}.  In  the determination of the
theoretical error, we use a correlation of $-0.8$ between the
\lbar\ and \lone. The resulting systematic uncertainty is $17.5\%$.

\subsection{Extraction of $|V_{ub}|$}
\label{sec:vub}
Combining the  measured ratio $\rusl$ with  the inclusive semileptonic
branching fraction of $\BR(\Bxlnu) = (10.87  \pm 0.18(stat) \pm
0.30(syst))\%$ as determined by \babar~\cite{Aubert:2002uf}, we obtain 

\begin{equation}
\BR(\Bxulnu) = (\allbr
\pm\allbrE(stat)
\pm\allbrEsyst(syst)
\pm\allbrEthHi(\lbar, \lone))\times10^{-3},
\end{equation}

\noindent where we have included the statistical and systematic
error of the branching fraction  $\BR(\Bxlnu)$. In the context of the OPE,
this result translates to

\begin{equation}
\Vub = (\allvub
\pm\allvubE(stat)
\pm\allvubEsyst(syst)
\pm\allvubEthHi(\lbar, \lone)
\pm\allvubEpert(pert)
\pm\allvubEhomb(1/m^3_b))\times 10^{-3}.
\end{equation}

\noindent where we have used the average $B$ lifetime of $\tau_B =
1.608 \pm 0.016 \ps$~\cite{pdg2002}.

The first error is the statistical uncertainty, with contributions
from data and Monte Carlo, the second one refers to the
experimental systematic uncertainty, the third one gives the
theoretical uncertainty in the efficiency determination and
extrapolation to the full \mX\ range, and the remaining errors refer
to uncertainties in the extraction of $|V_{ub}|$ from the branching
ratio.

\section{Measurement of the \BztoDslnu Branching Ratio }

\subsection{Event selection and reconstruction}
\label{sec:Analysis}

We select events containing a fully-reconstructed \Dstarm and an identified
oppositely-charged electron or muon.
The $\Dstarm\ellp$ pair is then required to pass kinematic cuts that
enhance the \BztoDslnu signal. Several control samples are used to characterize most of the \bkgs.
We define the following classification of the sources of signal and
backgrounds:
\begin{enumerate}
\item[1.] events with a correctly reconstructed \Dstar candidate:\\
   (a) events that originate from \BB events  
            with a correctly identified lepton candidate. 
	    They can be separated as:
            {\bf signal}
             ($\Bz\ra\Dstarm\ellp\nul(n\g)$ decays),
	    {\bf correlated-lepton background} 
             (\Bz or $\Bpm\ra\Dstarm X$, $X\ra\ellp Y$),
	    {\bf uncorrelated-lepton background} 
             ($\B\ra\Dstarm X$ and other $\B\ra\ellp Y$), 
	    {\bf charged \boldmath{\B} background}  
             ($\Bp\ra\Dstarm\ellp\nul X$),
       	    and {\bf neutral \boldmath{\B} background} 
             ($\Bz\ra\Dstarm\ellp\nul X$),\\
  (b) {\bf fake-lepton background}, e.g. 
              \BB events, with a hadron 
              misidentified as a lepton;\\
  (c) {\bf continuum background}, e.g. 
            $\ccbar\ra\Dstarm\ellp X$;
\item[2.] {\bf combinatorial-\boldmath{\Dstar} background}: 
          events with a mis-reconstructed \Dstarm candidate.
\end{enumerate}

\noindent
{\bf Lepton candidates}\\
Electron and muon candidates are selected according to particle identification criteria
which are described in detail elsewhere~\cite{babarnim}.
In addition to those criteria, they are required to have momentum greater than 
1.2~\gevc in the \FourS rest frame, since this cut significantly 
reduces the {\em correlated-lepton} background.
A sample enriched in {\em fake-lepton} background is also selected, where 
$\Dstarm h^+$ candidates are accepted if the charged track $h^+$ fails 
both electron and muon selection criteria looser than those required for 
lepton candidates, but fulfills all the other requirements.
This sample is used to determine the fraction of the {\em fake-lepton} 
background.\\
\\
\noindent
{\bf $D^{*-}$ candidates}\\
\Dstarm candidates are selected in the decay mode $\Dstarm\ra\Dzb\pim$. 
The \Dzb candidate is reconstructed in the modes
$\Kp\pim$, $\Kp\pim\pip\pim$, $\Kp\pim\piz$, and $\KS\pipi$.
The daughters of the \Dzb decay are selected according to the following
definitions: $\piz$ candidates are reconstructed from two photons with energy
greater than 30~\mev each, an invariant mass between 119.2 and
150.0~\mevcc, and a total energy greater than 200~\mev.
The mass of the photon pair is constrained to the $\piz$ mass and 
the photon pair is kept as a $\piz$ candidate if the \chisq
probability of the fit is greater than 1\%.
\KS candidates are reconstructed from a pair of charged particles with 
invariant mass within 15~\mevcc of the \KS mass.
The pair of tracks is retained as a \KS candidate if the \chisq
probability that the two tracks form a common vertex is greater than 1\%.
Charged kaon candidates satisfy loose kaon criteria for the
$\Kp\pim$ mode and tighter criteria for the $\Kp\pim\pip\pim$  and
$\Kp\pim\piz$ modes. 
For the $\Kp\pim\piz$ and $\KS\pipi$ modes, the candidate is retained if the 
square of the decay amplitude in the Dalitz plot for
the three-body candidate, based on measured amplitudes and 
phases~\cite{ref:E687}, is greater than 10\% of its maximum value across the 
Dalitz plot. 

\Dzb candidates are accepted if they 
have an invariant mass within 17~\mevcc of the \Dzb mass
in the $\Kp\pim$, $\Kp\pim\pip\pim$ and $\KS\pipi$ modes, and 34~\mevcc for the 
$\Kp\pim\piz$ mode.
The invariant mass of the daughters is constrained to the \Dzb mass and
the tracks are constrained to a common vertex in a simultaneous fit.  
The \Dzb candidate is retained if the \chisq probability of the fit
is greater than 0.1\%.

The low-momentum pion candidates for the $\Dstarm\ra\Dzb\pim$ 
decay are selected with total momentum in the laboratory frame less than 
450~\mevc, and momentum transverse to the beam line in the laboratory frame 
greater than 50~\mevc.
The momentum of the \Dstarm candidate in the \FourS rest frame 
is required to be between 0.5 and 2.5~\gevc. \\
\\
\noindent
{\bf $D^{*-}\ell^+$ candidates}\\
$\Dstarm\ellp$ candidates are selected if 
$|\cos\Delta\theta^*_{\rm thrust}| < 0.85$, where $\Delta\theta^*_{\rm thrust}$ 
is the angle between the thrust axis of the $\Dstarm\ellp$ candidate and the 
thrust axis of the remaining charged and neutral particles in the event. 
The distribution of $|\cos\Delta\theta^*_{\rm thrust}|$ is peaked at 1 for 
jet-like continuum events, and is flat for $B\overline B$ events, where 
the two $B$ mesons decay independently and practically at rest with 
respect to each other.
$\Dstarm\ellp$ candidates are selected if the \chisq probability of the fit to 
a common vertex of the lepton, of the $\pim$, and of the \Dzb candidates 
is greater than~1\%.

We define two angular quantities for each $\Dstarm\ellp$ candidate.
The first angle is $\theta_{\Dstar,\lepton}$, the angle between the \Dstarm 
and lepton in the \FourS rest frame.  
The signal distribution  of $\cos\theta_{\Dstar,\lepton}$ is peaked around 
$-1$ 
since the signal is more likely back-to-back in the \FourS rest frame, while
the distribution for the {\em uncorrelated-lepton} \bkg\ is flat.
A sample enhanced in \BztoDslnu signal events (called the {\em opposite-side} 
sample) is selected by requiring $\ctl<0$ while a  background control sample, 
representative of the {\em uncorrelated-lepton} background ({\em same-side} sample), 
is composed of  $\Dstarm\lepton$ candidates satisfying $\ctl\geq 0$
 
The second angle is $\theta_{\Bz,\Dsl}$, 
the inferred angle between the direction of the
\Bz and the vector sum of the \Dstarm and lepton candidate momenta in the \FourS rest frame. 
Assuming that the only missing particle in the \B 
reconstruction is a neutrino, the cosine of  $\theta_{\Bz,\Dsl}$ is calculated using:
 \begin{equation}
\cBDl= \frac{-(m^2_{\Bz} + m^2_{\Dsl}- 2\,E_{\Bz}^{*} E_{\Dsl}^{*})}
{2\,|\vec{p}_{\Bz}^{*}| |\vec{p}_{\Dsl}^{*}|   }\, .
 \label{eq:costhby}
 \end{equation}

\noindent 
All quantities in Eq.~\ref{eq:costhby} are defined in the \FourS rest frame.
The energy and the magnitude of the momentum of the \B are 
calculated from the \epem
center-of-mass energy and the \Bz mass.
For true \BztoDslnu events, \cBDl lies in the physical
region $[-1, +1]$, except for detector resolution and beam energy uncertainty.
Backgrounds lie inside and outside the range $[-1, +1]$.
In particular we use this variable to extract the 
$\B\ra\Dstarm\ellp\nul\X$ \bkg.   
Due to missing particles not accounted for when evaluating \cBDl for such a
\bkg\ event, the result will be less than the true value and it
can fall well below the physical region. 
We use this distribution to perform a \chisq binned fit for the 
$\B\ra\Dstarm\ellp\nul X$ background fraction. 

\subsection{Signal extraction}
\label{sec:bkg}

The candidates that pass the above selection criteria 
are distributed over two signal samples 
and ten background control samples
defined by the following characteristics:
\begin{enumerate}
\item whether the data were recorded {\em on} or {\em off} the \FourS resonance; 
\item whether the candidate lepton is {\em same-side} or {\em opposite-side}
to the \Dstarm
candidate; 
\item whether the lepton candidate passes the criteria for an electron, a
muon, or a fake lepton. 
\end{enumerate}
\noindent
The ten background samples are used to determine the background levels contaminating 
the two signal samples.

The {\em combinatorial}-\Dstar background 
can be distinguished from events with a \Dstarm by means 
of the mass difference $\deltam = m_{\Dstar} -m_{\Dz}$.
Each of the 12 samples described above 
is divided into 6 subsamples
according to the following characteristics that affect the \deltam
distributions.
\begin{enumerate}
\item the $\pim$ from the \Dstarm decay is reconstructed in the SVT only, 
or in the SVT and DCH (two choices):  
the \deltam resolution is worse when the $\pim$ is
reconstructed only in the SVT.
\item the \Dzb candidate is reconstructed in the mode $\Kp\pim$ or $\Kp\pim\piz$ 
or ($\Kp\pim\pip\pim$ or $\KS\pipi$) (three choices): 
the level of contamination
from {\em combinatorial}-\Dstar background and  the \deltam resolution depend on
the \Dzb decay mode. 
\end{enumerate} 

We fit the \deltam distributions for each of the total 72 subsamples.
The shape of the peak due to real \Dstarm decays is modeled 
by the sum of two Gaussian distributions; 
the mean and the variance of the Gaussian distributions and the relative 
normalization are free parameters. The shape of the {\em combinatorial}-\Dstar background is modeled with 
 \begin{equation}
 {1\over N} \left[ 1-\exp\left(-{\deltam - m_{\pim} \over c_1}\right)\right] 
  \left({\deltam\over m_{\pim} }\right)^{c_2},
 \label{eq:combback}
 \end{equation} 
where $N$ is a normalization constant,
$m_{\pim}$ is the mass of the $\pim$, and $c_1$ and $c_2$ are free
parameters.
\begin{figure}[!t]
\begin{center}
\includegraphics[width=3.in]{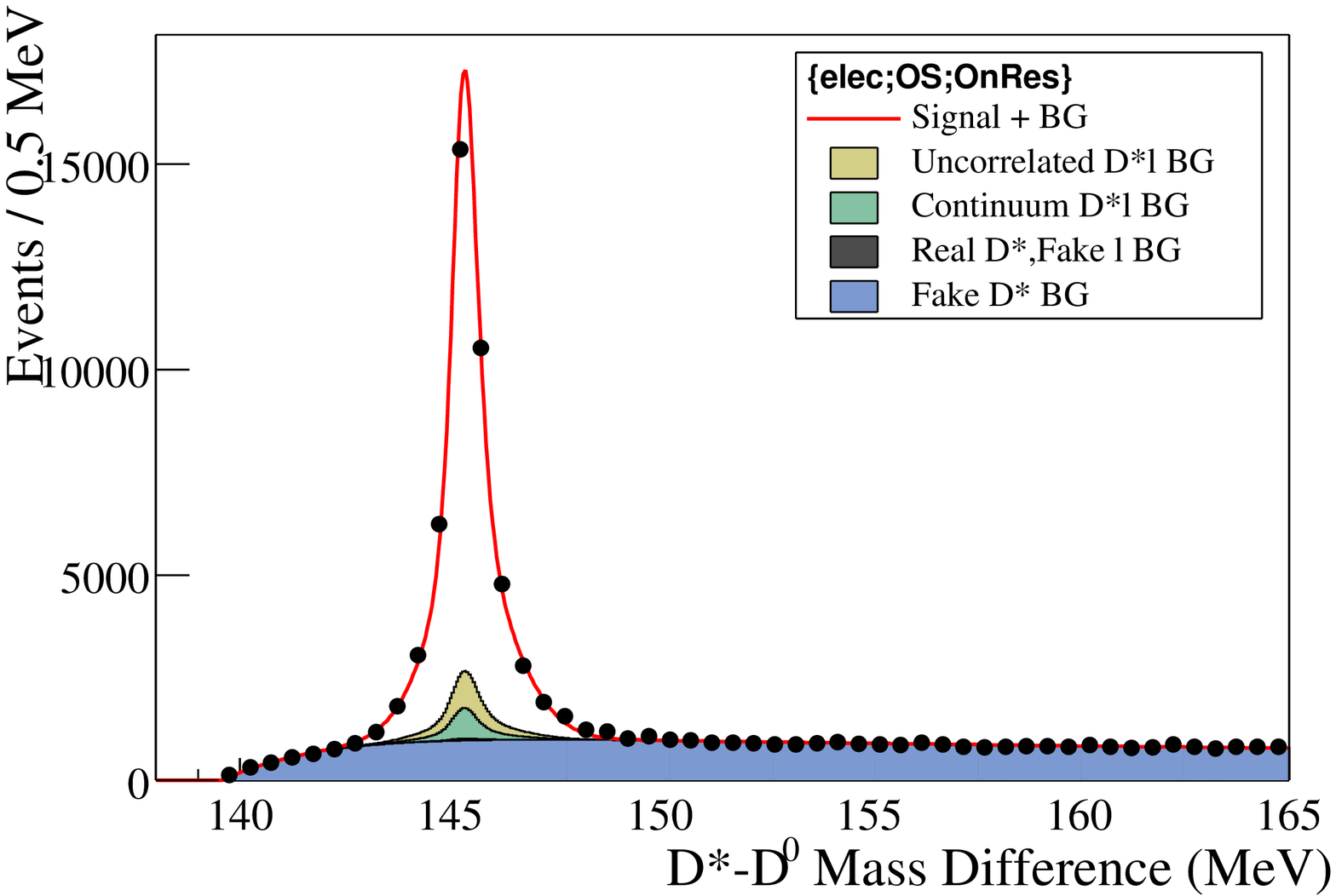}
\includegraphics[width=3.in]{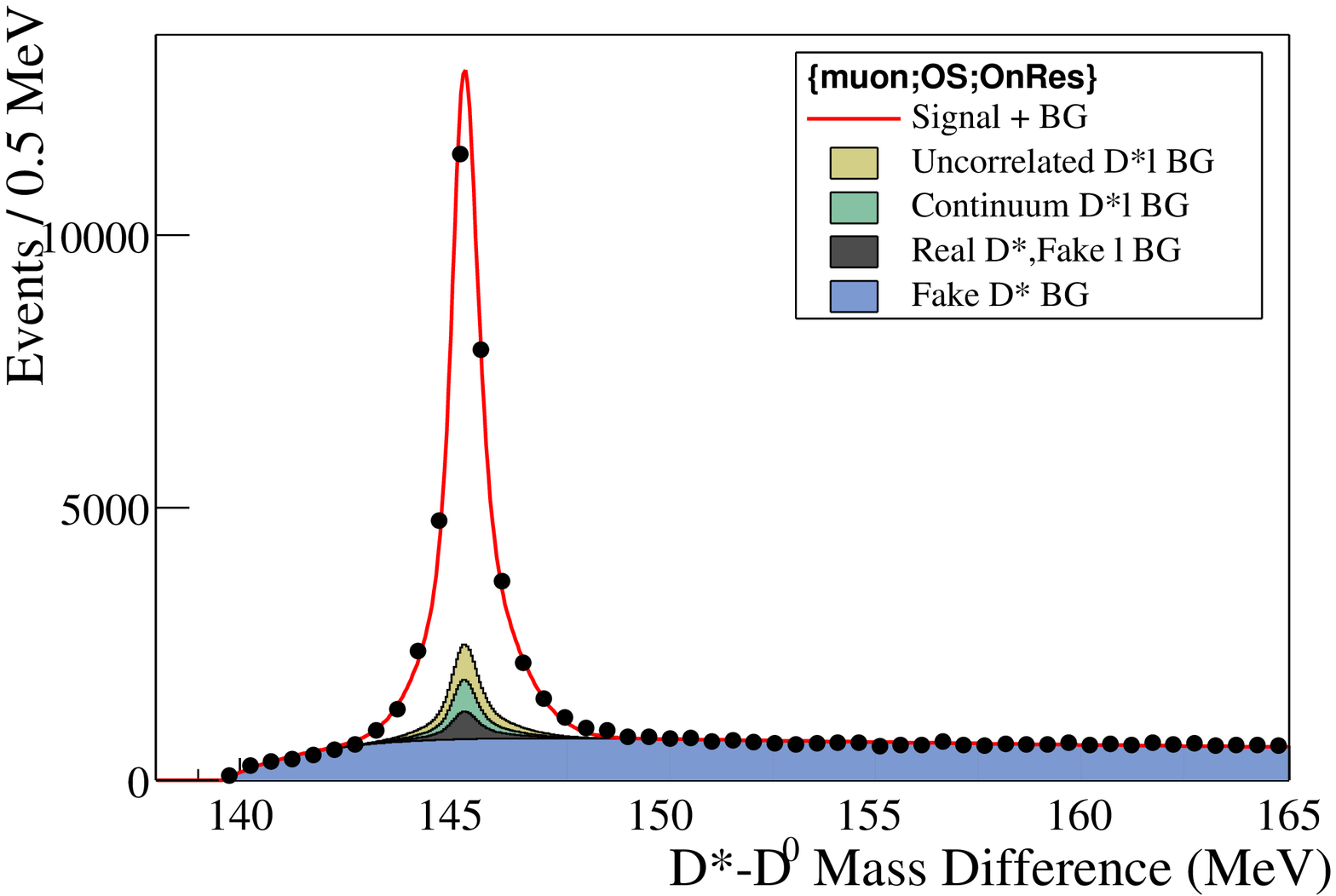}
\caption{\deltam distribution for events passing all selection criteria
for \BztoDslnu candidates, with (left) an electron or (right) a muon candidate.
The points correspond to the data.
The curve is the result of a simultaneous unbinned maximum likelihood fit to
this sample of events and a number of background control samples.
The {\em correlated-lepton} and \mbox{$\B\ra\Dstarm\ellp\nul\X$}
backgrounds are included in the signal distribution.}
\label{fig:massdifsig}
\end{center}
\vspace*{-0.5cm}
\end{figure}

We perform a simultaneous unbinned maximum likelihood fit to the 72 \deltam 
distributions to obtain the contributions of signal and background. 
The \deltam fits for the two signal samples 
({\em opposite-side} $\Dstarm\ep$ and $\Dstarm\mup$ candidates in 
{\em on-resonance} data) are shown in Fig.~\ref{fig:massdifsig}.  

\subsection{Determination of systematic errors}

One of the largest systematic errors, a relative 2.7\%, comes from 
uncertainties on \mbox{$\BR(\FourS\ra\BzBzb)$};
we use $\BR(\upsbpbm)/\BR(\upsbzbz) = 1.055\pm 0.055$\footnote{This value has 
been obtained correcting the average 
$\BR(\upsbpbm)/\BR(\upsbzbz)=1.072\pm 0.058$~\cite{pdg2002} for the 
lifetime ratio $\tau_{B^+}/\tau_{B^0}=1.083\pm0.017$~\cite{pdg2002}.}.
The data/Monte Carlo
tracking efficiency difference
(including both \piz and \KS identification efficiency,
but not the slow pion coming from $\Dstarm\ra\Dzb\pim$ decay) give a
2.7\% relative systematic error.
The uncertainties on $\BR (\Dzb)$~\cite{pdg2002} account for a relative 2.0\% error on 
\BR(\BztoDslnu). The systematic error associated with the uncertainty in the
relative composition of the $\B\ra\Dstarm\ellp\nul X$ \bkg\ accounts for a
relative 2.0\% error.
The systematic error coming from data/Monte Carlo reconstruction efficiency
differences for the slow pion coming from $\Dstarm\ra\Dzb\pim$ decay
accounts for a relative 1.9\%.
An estimate of the systematic error coming from 
the acceptance dependence on the HQET form factors 
has been evaluated by varying
the HQET Monte Carlo simulation input parameters within one standard
deviation of the measured values~\cite{ref:cleoformfactors}, giving
a relative error of 1.8\%. 
The systematic error coming from the uncertainty on the number of
\FourS events accounts for a relative 1.1\%.
All other systematic errors amount individually to less than a 1\% relative error.

\subsection{Results}

After combining the results for each of the eight signal channels
we obtain:
\[
\BR(\BztoDslnu)~=~(4.82 \pm 0.03 \pm 0.02 \pm 0.24 \pm 0.17)~\%.
\]
The first error comes from data statistics, the second from
Monte Carlo statistics, the third is the systematic error not related to PDG 
uncertainties, and the last one is the uncertainty due to the errors on 
PDG branching fractions for \FourS, \Dstarm, and \Dzb exclusive decays.

\section{Measurement of $\Delta m_d$ and $\tau_{B^0}$ with exclusively reconstructed $B^0 
\rightarrow D^{*-}l^+\nu_l$ decays}

This analysis \cite{dslnu} is based on a reduced sample of 23 million $B\bar{B}$ pairs recorded 
in the years 1999-2000 by BaBar. One of the two \B mesons is reconstructed in $B^0 
\rightarrow D^{*-}l^+\nu_l$ (using the same technique described in the previous chapter), 
and the charge of the final-state particles identifies the flavor of the $B$.
All the charged tracks in the event, except the reconstructed tracks from the $D^{*-}l^+$  pair, are used to identify the flavor of the other $B$ (referred to as $B_{tag}$). There are five types of tagging categories. The first two tagging categories rely on the presence
of a prompt lepton or charged kaons in the event, whose charge is correlated with the $b$-flavor of the decaying B.
The other three categories exploit a variety of inputs (e.g. slow pions, momentum of the track with the maximum center-of-mass momentum) with a neural network
technique.

The difference $\Delta t$ between the two $B$ decay times is determined from the measured separation $\Delta z = z_{D*l}-z_{tag}$ along the beam axis between the $D^{*-}l^+$ vertex and the $B_{tag}$ vertex.
The measured $\Delta z$ is converted into $\Delta t$ with the known $\Upsilon(4S)$ boost according to the relation $\Delta z = c \beta \gamma \Delta t$ which neglects the small $B$ momentum in the $\Upsilon(4S)$ frame. The resolution on the
$D^{*-}l^+$ vertex is about 70 $\mu m$ while the resolution on the $B_{tag}$
vertex is about 160 $\mu m$.

\begin{figure}[!t]
\begin{center}
\includegraphics[width=3.in]{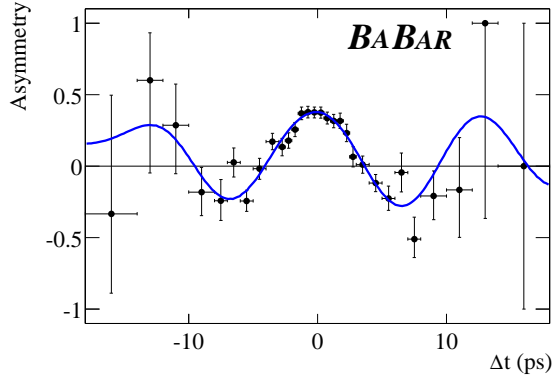}
\end{center}
\caption{The asymmetry plot for mixed and unmixed events in
an 80\%-pure signal sample
and the projection of the fit results. }
\label{fig:result-asym}
\end{figure}

The oscillation frequency $\Delta m_d$ and the lifetime $\tau_{B^0}$
are determined simultaneously with an unbinned maximum likelihood fit to the
measured $\Delta t $ distribution.
 Note that other mixing measurements fix $\tau_{B^0}$ to the world average and this is the source of the dominant systematic error. Also the resolution, the fraction of charged $B$, the mistag and the background parameters are floated in the fit. The results are: $\Delta m_d$ = $0.492 \pm 0.018(stat) \pm 0.013(syst)$ $ps^{-1}$ and $\tau_{B^0}$ = $1.523^{+0.024}_{-0.023}(stat) \pm 0.022(syst)$ $ps$. The correlation between $\Delta m_d$ and $\tau_{B^0}$ is -0.22. A correction is applied to both $\Delta m_d$ and $\tau_{B^0}$ which takes into account selection and fit biases. The uncertainty on such a correction is the
dominant systematic error.
Other BaBar measurements for $\Delta m_d$ and $\tau_{B^0}$ can be found in ~\cite{hadronic}, ~\cite{dilepton}, ~\cite{taupart} and ~\cite{tauhad}.
Figure~\ref{fig:result-asym} shows the mixing asymmetry defined as the difference between the number of unmixed and mixed events over their sum as a function of $\Delta t$.

\section{Conclusions}
We measured the CKM parameter $\Vub = (\allvub \pm\allvubE(stat) 
\pm\allvubEsyst(syst) \pm\allvubEthHi(\lbar, \lone) \pm\allvubEpert(pert) 
\pm\allvubEhomb(1/m^3_b))\times 10^{-3}$ using a novel technique based on 
the study of inclusive semileptonic decays on the recoil of fully 
reconstructed B mesons. This approach has a smaller systematic uncertainty 
and higher purity than previous measurements. We also measured the exclusive 
\BztoDslnu branching ratio $\BR(\BztoDslnu)~=~(4.82 \pm 0.03 \pm 0.02 
\pm 0.24 \pm 0.17)~\%$. This sample will allow us to extract the \Vcb 
parameter by studying its differential decay rate. \BztoDslnu events have 
been also used to measure simultaneously $\Delta m_d$ and $\tau_{B^0}$.

\end{document}